# Model Analysis: Assessing the Dynamics of Student Learning

Lei Bao
*Department of Physics,*
*The Ohio State University*
*174 West 18th Ave., Columbus, OH 43210-1106*

Edward F. Redish
*Department of Physics*
*University of Maryland*
*College Park, MD 20742-4111*

## Abstract

*In this paper we present a method of modeling and analysis that permits the extraction and quantitative display of detailed information about the effects of instruction on a class's knowledge. The method relies on a cognitive model of thinking and learning that represents student thinking in terms of patterns of association in long-term memory structures that we refer to as schemas or mental models. As shown by previous research, students frequently fail to recognize relevant conditions that lead to appropriate uses of their mental models and, as a result, can use multiple models inconsistently to treat problems that appear equivalent to an expert. Once the most common mental models have been determined via qualitative research, they can be mapped onto probing instruments such as a multiple-choice test. We have developed Model Analysis to analyze the results of these instruments that treats the student as if he/she were in a mixed state – a state which, when probed with a set of scenarios under diverse contextual settings, gives the probability that the student will choose a particular mental model to analyze the scenario. We illustrate the use of our method by analyzing results from the Force Concept Inventory, a research-based multiple-choice instrument developed to probe student's conceptual understanding of Newtonian Mechanics in a physics class. Model Analysis allows one to use qualitative research results to provide a framework for analyzing and interpreting the meaning of students' incorrect responses on a well-designed research-based multiple-choice test. These results can then be used to guide instruction, either for an individual teacher or for developers of reform curricula.*

## I. Introduction

### *The Problem: How can we evaluate what our students know in a large classroom?*

One of the most important things educational researchers have learned over the past few decades is that it is essential for instructors to understand what knowledge students bring into the classroom and how they respond to instruction. In small classes, this information can be obtained from careful one-on-one dialogs between student and teacher. In large classes, such as those typically offered in introductory science at colleges and universities, such dialogs are all but impossible. Instructors in these venues often resort to pre-post testing using their own or research-based closed-ended diagnostic instruments.

The results from these instruments tend to be used in a very limited way — through overall scores and average pre-post gains. This approach may ignore much valuable information, especially if the instrument has been designed on the basis of strong qualitative research, contains sub-clusters of questions probing similar issues, and has distracters that represent alternative modes of student reasoning.

In this paper, we present a method of *model analysis* that allows an instructor to extract specific information from a well-designed assessment instrument (test) on the state of a class's knowledge. The method is especially valuable in cases where qualitative research has documented that students enter a class with a small number of strong naïve conceptions that conflict with or encourage misinterpretations of the community-consensus view.

## *The Theoretical Frame: Knowing what we mean by "what our students know" requires a cognitive model.*

Although the desire to "understand what our students know" is an honorable one, we cannot make much progress until we both develop a good understanding of the characteristics of the system we are trying to influence (the student's knowledge structure) and have a language and theoretical frame with which to talk about it. Fortunately, much has been learned over the past few decades about how students think and learn and many theoretical models of human cognition have been developed.[1] Unfortunately, they have not yet coalesced into a single coherent model. Although significant differences among them remain, there is much common ground and much that is valuable for helping both the educational researcher and the teacher.

Despite the progress in cognitive science, most educational researchers analyzing real-world classrooms make little use of this knowledge. Many of the mathematical tools commonly used to extract information from educational observations rely on statistical methods that (often tacitly) assume that quantitative probes of student thinking measure a system in a unique "true" state. We believe that this model of thinking and learning is not the most appropriate one for analyzing a student's progress through goal-oriented instruction and is inconsistent with current models of cognition. (Examples will be given in the body of the paper.) As a result, the analysis of quantitative educational data can draw incomplete or incorrect conclusions even from large samples.

In this paper we hope to make a step towards ameliorating this situation. We begin by sketching a part of a theoretical framework based on the work of many cognitive researchers in education, psychology, and neuroscience. This framework describes a hierarchy of structures including elements of knowledge (both declarative and procedural), patterns of association among them, and mappings between these elements and the external world.[2] The well-known context-dependence of the cognitive response is represented by probabilities in the associative links. Note that these probabilities do not represent sampling probabilities associated with a statistical analysis of educational data taken from many students. These probabilities are fundamental and must be considered as lying within the individual student. This is the critical issue and will be explained in more detail in the body of the paper.

To be able to discuss the cognitive issues clearly, we define a series of terms and state a few fundamental principles. The most useful structure in our framework is the *mental model* — a robust and coherent knowledge element or strongly associated set of knowledge elements. We use this term in a broad and inclusive sense.[3] A mental model may be simple or complex, correct or incorrect, recalled as a whole or generated spontaneously in response to a situation. The popular (and sometimes debated) term *misconception* can be viewed as reasoning involving mental models that have problematic elements for the student's creation of an expert view and that appear in a given population with significant probabilities (though not necessarily consistently in a given student). We stress that our use of this term implies no assumption about the structure or the cognitive mental creation of this response. In particular, we do not need to assume either that it is irreducible (has no component parts) or that it is stored rather than generated. Furthermore, we are not returning to the idea that students have consistent "alternative theories" of the world. The data on the inconsistency of student responses in situations an expert would consider equivalent is much too strong to ignore (Clough & Driver, 1986; Maloney & Siegler 1993; Wittmann 1998). Instead, we describe the state of our student in a more complex way.

## *Model Analysis: A cognitive approach to evaluation*

For a particular physics concept, questions designed with different context settings can activate a student to use different pieces of knowledge. In Model Analysis, we represent the student's mental state as a vector in a "model space" spanned by a set of basis vectors, each representing a unique type of student reasoning (referred to as a model) that has been identified through qualitative research. The vector representing a single student's state is often a linear combination of the basis vectors of the "model space". The coefficient of a particular basis vector in the student state vector are taken to be the (square root of the) probability that the student will activate that particular model when presented with an arbitrary scenario chosen from a set of physics scenarios that involve the same underlying concept but that have different contextual settings.

---

[1] See, for example, the overview (Thagard, 1996), which discusses many models and contains a large number of references.

[2] Meta-elements of judgment, evaluation, and epistemology are also of great importance but are not considered in this paper.

[3] Note that this term (and the other terms we define) appear frequently in the cognitive and educational literature, often in undefined and inconsistent ways. In order to try to establish some level of clarity and consistency in the global picture we are developing, we explicate our definitions in detail.

In the rest of the paper, we show that this theoretical frame can be applied to develop new methods of analysis of traditional probes of instruction and that this new analysis can yield more detailed insight into instruction than is available using standard methods.

## *An Example: Probing understanding of Newtonian Physics in university classrooms*

As an example of our approach, we develop a mathematical framework for describing the implications of our cognitive model in the context of a typical probe of a student's cognitive state in a large classroom — a carefully designed multiple-choice examination. As a result of the desire to probe large groups efficiently, modern instruments designed to probe students' conceptual knowledge are often written as multiple-choice tests. The best of these are developed as the result of extensive qualitative research into student thinking and the most common student misconceptions determined as a result of that research are activated by carefully written distracters. Two examples of such tests in physics are the Force Concept Inventory (FCI) and the Force-Motion Concept Evaluation (FMCE) (Hestenes *et al*., 1992; Thornton & Sokoloff, 1998). These two tests are appropriate for probing aspects of students' conceptual understanding of high school or introductory college treatments of Newtonian mechanics. They are widely given as pre- and post-tests and have played an important role in making physics teachers aware of the limited effectiveness of traditional methods of instruction (Mazur, 1997).

Most analyses of the results from the FCI and FMCE compare the pre- and post-test scores of a class and measure an overall "efficiency of instruction" by calculating the fraction of the possible gain, $g$, attained by the class (Hake, 1998). The efficiency $g$ is given by

$$g = \frac{PretestPercentage - PosttestPercentage}{1 - PretestPercentage}$$

Unfortunately, while giving a global overview of teaching effectiveness, such a result blends together a variety of distinct learning issues and makes it difficult for an instructor to draw any detailed conclusions about what in his/her instruction was effective or ineffective. This limits the utility of such tests for providing specific guidance to a teacher/researcher for the reform of instruction. In educational statistics, researchers employ advanced methods such as factor analysis to extract possible latent model-like traits (factors) that underlie students' responses. However, most of these methods assume consistency in students' activation and application of conceptual knowledge and often rely on score-based data, both of which can lead to difficulty in extracting explicit information on student conceptual models. For example, a factor analysis of FCI results leads to the conclusion that there are no distinct factors, other than the obvious cluster that refers to the conceptually distinct Newton's 3rd Law (Huffman & Heller, 1995).

Our approach assumes that the most commonly used mental models are identified through extensive qualitative research. These known factors can then be mapped onto the choices of an appropriately designed multiple-choice test. Since the mental states of the individual students tend to be mixed, especially when they are making a transition from an initial state dominated by a naïve incorrect model to an expert state, inconsistent responses are interpreted not as an absence of factors but as a measure of the degree of mixture in the student's state.

Our analysis method is mathematically straightforward and can be easily carried out on a standard spreadsheet. The result is a more detailed picture of the effectiveness of instruction in a class than is available with analyses of results that do not consider the implications of the incorrect responses chosen by the students.

*Map of the Paper*

The paper is structured as follows. In section II, we introduce the theoretical structure we use as a framework for describing student learning. In section III, we define a mathematical representation, the model space, to study the dynamics of student models. In section IV, we introduce an algorithm, model estimation, to analyze student models with multiple-choice questions. Section V gives a specific example in which we apply model estimation to FCI data. In section VI, we compare our model analysis with factor analysis and discuss issues in measurement technology. We conclude with a summary in section VII.

## II. The Theoretical Frame: A Model of Cognition

The fundamental model of thinking and learning we will rely on has been built through the interaction of three kinds of scientific research:

- *Ecological*: phenomenological observations of normal behavior, especially in educational environments, (mostly carried out by educational researchers) (Bransford, Brown, & Cocking, 1999; McDermott & Redish, 1999)

- *Psychological*: careful studies of responses to highly simplified experiments designed to get at fundamental cognitive structures, (mostly carried out by psychologists) (Anderson, 1995; Baddeley, 1988) and

- *Neurological*: studies of the structure and functioning of the brain (mostly carried out by neuroscientists) (Shallice, 1988; Squire & Kandel, 1999; Fuster, 1999).[4]

In each of these areas of research, numerous models of cognition have been built. Although there is still much uncertainty about what the final model will look like, there is beginning to be a significant overlap. We particularly rely on elements that have significant support in all three areas.

*Memory*

The critical issue for teaching and learning is memory. A partial map of the structure of memory that has been developed by cognitive and neural scientists is sketched in figure 1. This model is rich in implications for understanding teaching and learning.[5] For the purpose of this paper, a few fairly simple ideas suffice.

There are two important functional components to memory:

- working memory, a fast storage of small size and short (few seconds) decay time, and
- long-term memory, a slow storage of very large capacity that may last for many years.

Much is known about this system, but at present there are also many contentious issues and uncertainties. Although what has been learned about working memory and about the creation of long-term memory has powerful implications for the details of instructional delivery, we focus here on the structural characteristics of long-term memory since, for this paper, we are interested in evaluating the success of instruction.

A few principles briefly describe some characteristics of long-term memory that can help us better understand the responses of students.

1. *Long-term memory can exist in (at least) 3 stages of activation: inactive, primed (ready for use), and active (immediately accessible).[6]*

---

[4] The neurological studies probe the cognitive process through modern non-invasive probes of brain function during cognitive tests, such as fMRI (functional magnetic resonance imagery) and PET (positron emission tomography), and through studies of the effects of brain injury. Important albeit more indirect evidence has also been obtained from invasive experiments with animals ranging in complexity from Aplysia (a sea-slug) to non-human primates.

[5] The components of this model and the evidence for them are discussed in (Baddeley, 1989), (Fuster, 1999), (Squire & Kandel, 1999), and Shallice (1989).

2. *Memory is associative and productive. Activating one element leads, with some probability, to the activation of associated elements. The activation of a memory often contains data, reasoning elements, and mappings of the memory elements onto input structures.*
3. *Activation and association are context dependent. What is activated and subsequent activations depend on the context, both external and internal (other activated elements).*

These principles are supported by a wide variety of studies ranging from the ecological to the neurological. Many complex and detailed cognitive models have been built that embody these principles (Thagard, 1996).

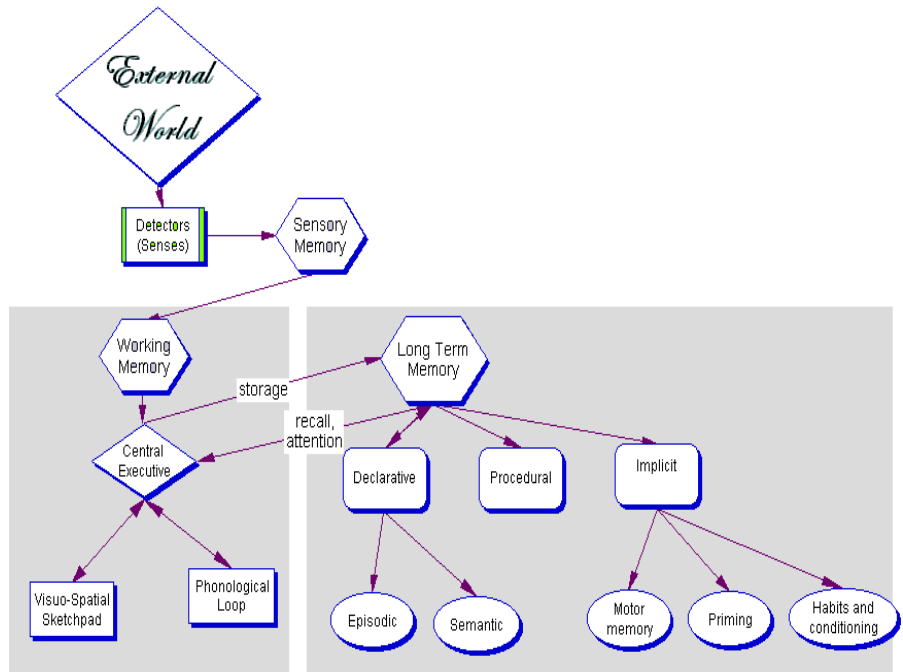

*Figure 1. A partial model of structures in memory. Adapted from (Baddely, 1988) and (Squire and Kandel, 1999).*

## *Context Dependence*

Some classical approaches in psychology and cognitive science treat learning as highly general processes and provide less emphasis on the context or situation in which the learning takes place. In contrast, situated cognition has its main focus on the interactions between people and the historically and culturally constituted contexts (Lave, 1991; Rogoff, 1991; Resnick, 1991; Greeno, 1992; Clancey, 1993). Learning is understood largely in terms of the achievement of appropriate patterns of behavior, which are strongly tied to and sustained by the contexts. Between these extremes, a considerable number of recent studies have emphasized understanding the involvement of contexts in learning while keeping the consideration of the constraints associated with abstract high-level mental constructs. Vosniadou (1994) developed the idea of mental models to account for the functional mental elements that subjects create in specific contexts based on their fundamental beliefs which are abstract and less tied with contexts.

An example from the physics education literature illustrates the implications of the context dependence of recall from long term memory in problem solving. Steinberg & Sabella (1997) asked two equivalent questions on Newton's first law to students in engineering physics at the University of Maryland. In both questions, the students were asked to compare the forces acting on an object moving vertically at a constant velocity. One question was phrased in physics terms using a laboratory example ("A metal sphere is resting on a platform that is being lowered smoothly at a constant velocity…"). The other was phrased in common speech using everyday experience ("An elevator is being lifted by a cable..."). In both problems, students were instructed to ignore friction and air resistance.[7]

---

[6] Indeed, (Fuster 1999) and others (Miyake & Shah 1999) suggest that working memory may not correspond to separate neurological structures from long-term memory but could correspond to a state of activation of the neural network.

On the physics-like problem, 90% of the students gave the correct answer that the normal force on the sphere is equal to the downward force due to gravity. On the everyday problem, only 54% chose the correct answer: the upward force on the elevator by the cables equals the downward force due to gravity. More than a third, 36%, chose the answer to this second problem reflecting a common incorrect model: the upward force on the elevator by the cables is greater than the downward force due to gravity.

Strong context dependence in student responses is very common, especially when students are just beginning to learn new material. Students are unsure of the conditions under which rules they have learned apply and they use them either too broadly or too narrowly. (Understanding how these conditions apply is a crucial part of learning a new concept.) Students often treat problems that look equivalent to an expert quite differently.

### *Structures in Long-Term Memory*

A great deal has been learned about the organization of long-term memory and reasoning in a variety of contexts including the interpretation of text (Gernsbacher 1994), the learning of grammar (Morgan 1986), the interpretation of mathematics (Schoenfeld 1985), and the interpretation of physical phenomena (diSessa and Sherrin 1998). Since our primary interest is science learning in general and the learning of physics in particular, we restrict our discussion here to thinking and learning about physical phenomena. Much of what has been learned under this rubric has analogs in other areas.

From the complex analysis of the structure of long-term memory carried out by many researchers, we identify five structures that are particularly relevant for the understanding of physical phenomena and for the study of physics:

- reasoning primitives,
- facets,
- schemas,
- mental models,
- physical models.

The first two items in the list are the basic elements out of which physical reasoning is built. These ideas come from those developed by diSessa (1993) and Minstrell (1992), but we make the cut between the structures they describe in a slightly different fashion in order to make use of the strengths of both descriptions.

DiSessa investigated people's *sense of physical mechanism*, that is, their understanding of why things work the way they do. What he found was that many students, even after instruction in physics, often come up with simple statements that describe the way they think things function in the real world. They often consider these statements to be "irreducible" – as the obvious or ultimate answer; that is, they can't give a "why" beyond it. "That's just the way things work," is a typical response. DiSessa refers to such statements as *phenomenological primitives* or *p-prims*. Minstrell observed that students' responses to questions about physical situations can frequently be classified in terms of reasonably simple and explicit statements about the physical world or the combination of such statements. Minstrell refers to such statements as *facets*.

Sometimes diSessa's p-prims are general in character and sometimes they refer to specific physical situations and are similar to Minstrell's facets. We restrict the term "primitive" to describe the finest-grained cognitive element in our hierarchy, the *reasoning primitive* — the abstract p-prims in diSessa's collection and the abstractions of those that refer to more specific physical situations. We use the term *facet* to refer to the mapping of a reasoning primitive into a physical situation. Note that we do not make any hypothesis about how a facet is actually structured neurologically. diSessa calls some elements (that we refer to as facets) p-prims because they are irreducible to the user. Be this as it may, we find it useful to separate the concepts along the lines described.

---

[7] In his dissertation (Sabella, 1999), Sabella demonstrated that the results described in (Steinberg & Sabella, 1997) did not strongly depend on which case was taken as rising or falling.

An example of where we would separate differently from diSessa is in the p-prim "continuous push" — a constant effort is needed to maintain motion. We would refer to this as a facet which is a mapping of the reasoning primitive: a continued cause is needed to maintain a continued effect. In his long paper on the subject, diSessa recognizes these different levels of abstraction but chooses not to apply a different terminology to them. We suspect our choice of terminology will be more useful in helping to design instructional environments to help students re-map their facets, but at this point our evidence is only anecdotal.

The critical items in our hierarchy are *mental model* and *schema*. Both of these terms have been used in the literature in a variety of different ways and are often undefined. We use the term schema to mean a cognitive element or set of cognitive elements that activate together in a cascading chain of activation or priming in response to a stimulus or situation presented to the student. This is our most general term and refers to any repeatable pattern of association of mental elements. We reserve the term *mental model* for a robust and reasonably coherent schema. Our use of these terms are very broad and meant to cover a wide variety of circumstances ranging from the cueing a single explanatory facet to the activation and priming of a coherent extensive and complex knowledge structure. There are certainly a variety of structures at this level and their exploration is currently providing work for many hands (Vosniadou 1994; diSessa & Sherin, 1997; Chi et al. 1994).

Finally, we introduce the term *physical model* to indicate a type of mental model that is commonly used by a particular population in working with situations related to a particular concept. Physical models are mental models that are explicitly organized around specific ontological assumptions: certain objects that are assumed exist and have specific characteristics, properties, and interactions. Properties and processes in such a model arise from combining the assumed ontological structures and properties.

A physical model may or may not agree with our current community consensus view of physics. For example, the phlogiston picture of thermodynamics was organized around an image of a physical substance that we now understand cannot be made consistent with physical observations, so this physical model does not agree with our current community physical model. Two individuals well versed in a topic may have different physical models of the same system. For example, a circuit designer may model electric circuits in terms of resistors, inductors, capacitors, and power sources – macroscopic objects with particular properties and behaviors. A condensed-matter physicist may think of the same system as emerging from a microscopic model of electrons and ions.

### *How should one view the context-dependence of mental elements and operations?*

The context dependence of the cognitive response may be considered in a variety of ways. From the point of view of the student, his/her mental system may feel perfectly consistent, despite appearing inconsistent to an expert. The student might use a mental model inappropriately because he/she has failed to attach appropriate conditions to its application (Reif & Allen, 1992), or the student might fail to associate a mental model with a circumstance in which it is appropriate, or students may associate to different mental models in equivalent circumstances because they cue on irrelevant elements of the situation and do not notice that the circumstances are equivalent.

From the point of view of the cognitive researcher, it may be of great interest to consider the student as always being in a consistent mental state or as flipping from one mental state to another in response to a variety of cues. However, from the point of view of the educational researcher or of the instructor interested in goal-oriented instruction — that is, in acculturating students to understand particular community-developed viewpoints — we suggest that there is considerable value in analyzing the student's thinking as projected against an expert view. The "expert" here needs to be both a subject expert and an expert in education research so as not to undervalue or misunderstand the view of the student. For example, in considering the motion of compact objects, a naïve physics student might view objects in terms of a unified concept of "motion" instead of separating the ideas of position and velocity. The mental models used by the student must be understood in terms of their own internal consistencies, not as "errors" when projected against the expert view.

Suppose we prepare a sequence of questions or situations in which an expert would use a single, coherent mental model. We refer to these as *a set of expert-equivalent questions*. Further, suppose, that when presented with some questions from such a set, a particular student can use a variety of mental models instead of a single coherent expert model. Such a situation is extremely common in many learning situations and is well documented to occur frequently in introductory physics (Maloney and Siegler, 1993; Wittmann, 1998). The elements of each question in the set that trigger a student to choose a particular model (or a set of models) depend, not only on the student's educational history, but even on the student's mental state at the particular instant the question is probed, in particular, on which mental elements are primed (Sabella, 1999). Since both the educational history and the student's mental state are difficult if not impossible to determine, we propose that the most appropriate way of treating this situation is probabilistically.

If a student always uses a particular mental model in a reasonably coherent way in response to a set of expert-equivalent questions we say they are in a *pure model state*. If the student uses a mixture of distinct mental models in response to the set of questions we say the student is in a *mixed model state*. We view the individual student who is in a mixed state as simultaneously occupying a number of distinct model states with different probabilities. The distribution pattern of the probabilities gives a representation for a mixed state.

When the student's state is probed by the presentation of a particular question or scenario, the student will often respond by activating a single mental model.[8] We view the student's mental state as having been *momentarily collapsed by the probe into the model state* selected.

Note that the statistical (probabilistic) character of the student state arises from the presentation of a large number of questions or scenarios, not from the probing of multiple students. We view the context dependence of mental model generation as a fundamental statistical character built into the individual. The statistical treatment is a way of treating many "hidden variables" in the problem that are both uncontrollable and possibly unmeasurable even in principle.

This approach, which will be developed mathematically below, provides an alternative assumption to the one traditionally made (Spearman, 190x), that a probe of the state of a student yields the "true value" plus some random error:

$$M = T + X.$$

We propose that a more appropriate model for analyzing student thinking is to consider the distribution of a student's inconsistent results on a set of equivalent questions as a measure of a property of the student, not as "random error." Of course random errors do occur. In this paper, the probabilistic distribution is interpreted as fundamental and representing the characteristics of students. In future papers we will consider the implication of random error and how to treat sampling issues in the environment we describe.

As discussed earlier, mental models are productive structures that can be applied to a variety of different physical contexts to generate explanatory results. Mental models can be either complex or simple. For this work, to clarify the nature of our model and method, we have chosen to restrict our considerations to simple

---

[8] The process by which this selection is made can be quite complex. In some cases, only a single model is activated. In others, multiple models are activated and an "executive process" is assumed to make a choice of one, suppressing other models. When such a choice is difficult to make, a student can get into an explicit state of confusion where several models appear to be equally plausible (but generating contradictory results) and the student can't determine which one is more appropriate to use. Depending on the design of the probing instruments, such states may or may not be extracted. For example, multiple-choice single response questions often force students to pick one answer and thus can only measure the existence of one of the models, while multiple-choice multiple-response questions can extract information about such mixing states. Although this topic has not explicitly been studied (to our knowledge) in physics education, there is extensive research on the issue in cognitive and neuroscience. (See for example, Shallice & Burgess, 1998.)

models — essentially single facets. We do not intend to imply that all student reasoning is describable by such a simple situation. There are, however, numerous examples of such situations, and we intend to demonstrate the value of our approach by applying it to this simple, highly restrictive situation.

The use of mixed models or competing concepts appears to be a typical and important stage in student learning of physics (Bao, 1999; Bao, Hogg, & Zollman, 2001; Thornton & Sokoloff, 1998; Maloney & Siegler, 1993). To study the dynamical process of students' applying their models, we first define two important concepts: *common models* and *student model states*.

**Common Models**

When the learning of a particular physics concept is explored through systematic qualitative research,[9] researchers are often able to identify a small, finite set of commonly recognized models. (Marton 1986) These models often consist of one correct expert model and a few incorrect or partially correct student models. Note that different populations of students may have different sets of models that are activated by the presentation of a new situation or problem. When presented with novel situations, students can activate a previously well-formed model or, when no existing models are appropriate, they can also create a model on the spot using a mapping of a reasoning primitive or by association to salient (but possibly irrelevant) features in the problem's presentation. The identified common student models can be formed in both ways. Although the actual process is not significant in the research of this paper, the specific structure of the models involved may have important implications for the design of instruction.

**Student Model State**

When a student is presented with a set of questions related to a single physics concept (a set of expert equivalent questions), two situations commonly occur.

1. The student consistently uses one of the common models in to answer all questions.

2. The student uses different common models and is inconsistent in using them, i.e., the student can use one of the common models on some expert-equivalent questions and a different common model on other questions.

The different situations of the student's use of models are described as student *model states*. The first case corresponds to a pure model state and the second case to a mixed model state.

When analyzing the use of common models, it is necessary to allow an additional dimension to include other less common and/or irrelevant ideas that student might come up with. To collect this set of responses we identify a *null* model — one not describable by a well-understood common model. With the null model included, the set of models becomes a complete set, i.e., any student response can be categorized.[10] Specific examples of common models and student model states will be discussed in later sections.

If a set of questions has been carefully designed to probe a particular concept, we can measure the probability for a single student to activate the different common models in response to these questions. We can use these probabilities to represent student model state. Thus, a student's model state can be represented by a specific configuration of the probabilities for the student to use different physical models in a given set of situations related to a particular concept.

Figure 2 shows a schematic of the process of cueing and activating a student's model, where $M_1 \ldots M_w$ represent the different common models (assuming a total of $w$ common models including a null model), and $q_1 \ldots q_w$ represent the probabilities that a particular situation will result in a student activating the corresponding model.[11] For convenience, we consistently define $M_1$ to be the expert model and $M_w$ to be the null model. The possible incorrect models are represented with $M_2 \ldots M_{w-1}$.

---

[9] These researches should always involve detailed individual student interviews and the results should also be verifiable by other researchers.

[10] Of course, in addition to collecting random and incoherent student responses, coherent models that have not yet been understood as coherent by researchers may well be classified initially as "null". When a significant fraction of student responses on a particular question winds up being classified as null, it is possible that a better understanding of the range of student responses needs to be developed through qualitative research.

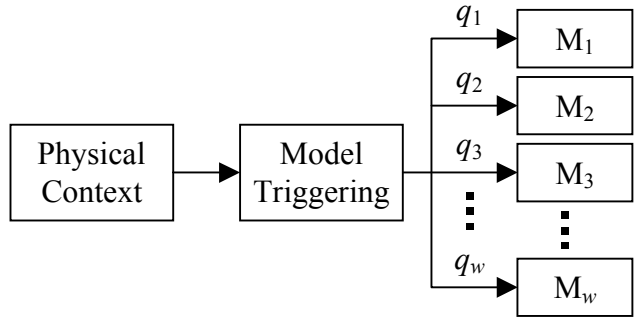

*Figure 2. Using a set of questions designed for a particular physics concept, we can measure the probability for a single student to use different physical models in solving these problems. In the figure, $M_1 \ldots M_w$ represent the different physical models (there are a total of w physical models including a null model), and $q_1 \ldots q_w$ represent the probabilities for a student being triggered into activating the corresponding models.*

## III. The Student Model Space: A Mathematical Representation

We represent the state of the student with respect to a set of common models by a linear vector space. Each model is associated with an element of an orthonormal basis, $\mathbf{e}_\eta$:

$$\mathbf{e}_1 = \begin{pmatrix} 1 \\ 0 \\ \vdots \\ 0 \end{pmatrix}; \quad \mathbf{e}_2 = \begin{pmatrix} 0 \\ 1 \\ \vdots \\ 0 \end{pmatrix}; \quad \ldots \quad \mathbf{e}_w = \begin{pmatrix} 0 \\ 0 \\ \vdots \\ 1 \end{pmatrix} \tag{1}$$

where $w$ is the total number of physical models being considered (including a null model) associated with the concept being probed.

We refer to the space spanned by these model vectors as the *model space*. As discussed in section III, in general, the student can be expected to be in a mixed model state. For a given instrument, we represent this state using the probabilities for a student to be cued into using each of the different models. In principle, these probabilities can be probed in experiments; however, a precise determination is often difficult to achieve even with extensive interviews. But, in practice we can obtain estimations of this probability with properly designed measurement instruments.

A convenient instrument is a set of research-based multiple-choice questions. Suppose we give a population of students $m$ multiple-choice single-response (MCSR) questions on a single concept for which this population uses $w$ common models. Define $\vec{Q}_k$ as the $k^{\text{th}}$ student's probability distribution vector measured with the $m$ questions. Then we can write

$$\vec{Q}_k = \begin{pmatrix} q_1^k \\ q_2^k \\ \vdots \\ q_w^k \end{pmatrix} = \frac{1}{m} \begin{pmatrix} n_1^k \\ n_2^k \\ \vdots \\ n_w^k \end{pmatrix} \tag{2}$$

---

[11] Note that given different sets of questions, the measured probabilities can be different. The measured student model state is a result of the interaction between the individual student and the instrument used in the measurement and should not be taken as a property of the student alone. This is discussed in detail in the next section.

where $q_\eta^k$ represents the probability for the $k^{th}$ student to use the $\eta^{th}$ model in solving these questions and $n_\eta^k$ represents the number of questions in which the $k^{th}$ student applied the $\eta^{th}$ common model. We also have

$$\sum_{\eta=1}^{w} n_\eta^k = m. \tag{3}$$

In eq. (2) we have taken the probability that the $k^{th}$ student is in the $\eta^{th}$ model state to be $q_\eta^k = n_\eta^k/m$. Note that $q_\eta^k$ is affected by the specific question set chosen. There is not a unique state of the student for a given concept and set of mental models. The student model state represents an interaction between the student and the particular instrument chosen.

To see why this is the case, consider an infinite set of expert equivalent questions concerning a particular concept that an individual student might consider as requiring two different models, model A or model B, depending on the presence of absence of a particular (actually irrelevant) element in the problem. Assume that if the element is present, the student strongly tends to choose model A, otherwise they will choose model B. Since the set of questions can contain infinitely many items having the element and infinitely many items that do not, the instrument designer may create an instrument that has any proportion of the items containing the irrelevant element. The percentage of student choices of models A or B will depend on the number of items on the test containing A.

The student state as measured by a particular instrument therefore depends on both the student and the instrument. Since we are concerned with evaluating normative instruction, in which the student is being taught a particular model or set of models, the choice of the proportion of questions depends on normative goals — what the instrument designer considers it important for the student to know. *The student state should therefore be thought of as a projection of student knowledge against a set of normative instructional goals, not as an abstract property belonging to the student alone.*[12] For the purpose of assessment, researchers can develop (through systematic research on student models) a rather standardized set of questions based on the normative goals. These questions can then be used to provide comparative evaluation on situations of student models for different populations.

We do not choose to consider the set of probabilities, $\vec{Q}_k$, to represent the model state of the $k^{th}$ student. Rather, we choose to associate the student state with a vector consisting of the *square root* of the probabilities. We refer to these square roots as the *probability amplitude*. In principle, either approach might be considered. In practice, there are considerable advantages to the square root choice, as it naturally leads to a convenient structure, the density matrix, as we will see below. We therefore choose to represent the model state for the $k^{th}$ student in a class with a vector of unit length in the model space, $\mathbf{u}_k$:

$$|\mathbf{u}_k\rangle = \begin{pmatrix} \sqrt{q_1^k} \\ \sqrt{q_2^k} \\ \vdots \\ \sqrt{q_w^k} \end{pmatrix} = \frac{1}{\sqrt{m}} \begin{pmatrix} \sqrt{n_1^k} \\ \sqrt{n_2^k} \\ \vdots \\ \sqrt{n_w^k} \end{pmatrix} \tag{4}$$

where

$$\langle \mathbf{u}_k | \mathbf{u}_k \rangle = \sum_{\eta=1}^{w} q_\eta^k = 1 \tag{5}$$

We choose to represent the vector using the "bra-ket" notation in order to distinguish it from the probability vector, $\vec{Q}_k$. The ket state $|\mathbf{u}_k\rangle$ represents the column vector describing the state of the $k^{th}$ student in the model space. The bra state $\langle \mathbf{u}_k |$ is a row vector, the transpose of the ket state. By the standard rules

---

[12] Methods of assigning instrument-independent model-probability descriptions to students will be considered in a subsequent paper. See (Bao, 2001).

*11*

of matrix multiplication, the two put together in "bra-ket" form creates a bracket, which is a number, the dot product of the two vectors.[13]

## IV. Analyzing Student Models with Multiple-Choice Instruments

Using our mathematical representation, we can analyze student responses to multiple-choice questions to measure student model states and study the evolution of a class's learning. The development of an effective instrument should always begin with systematic investigations on student difficulties in understanding a particular concept. Such research often relies on detailed interviews to identify common models that students may form before, during and after instruction. Using the results from this research, multiple-choice questions can be developed where the choices of the questions is designed to probe the different common student models.[14] Then interviews are again used to confirm the face validity of the instrument, elaborate what can be learned from the data, and start the cyclic process to further develop the research.

In physics education, researchers have developed research-based multiple-choice instruments on a variety of topics. The two most popularly available instruments on concepts in Newtonian mechanics are the FCI and FMCE (Hestenes, Wells, & Swackhammer, 1992; Thornton, 1994). The questions were designed to probe critical conceptual knowledge and their distracters are chosen to activate common naïve conceptions. As a result, many of the questions on these tests are suitable for use with the model analysis method. In this paper, we use the data of the FCI test from engineering students in the calculus-based physics class at the University of Maryland. Results of the FMCE test with students from other schools are discussed in (Bao, 1999).

**The force–motion model**

An example in Newtonian mechanics where students commonly have a clearly defined and reasonably consistent facet is the relation of force and motion. Student understanding of the force – motion connection has been thoroughly studied for the past two decades and researchers have been able to develop a good understanding of the most common student models (Viennot, 1979; Watts, 1983; Champagne, Klopfer & Anderson, 1980; Clement, 1982; Galili & Bar, 1992; Halloun & Hestenes 1985). A commonly observed student difficulty is that students often think that a force is always needed to maintain the motion of an object. As a result, students often have the idea that there is always a force in the direction of motion. For the population in our introductory physics class, this is the most common incorrect student model related to the force–motion concept. Some even consider that the force is proportional to the velocity. In the physics community model, an unbalanced force is associated with a change in the velocity — an acceleration. Therefore, for this concept, we can define three common models:

Model 1:   An object can move with or without a net force in the direction of motion. (expert model)
Model 2:   There is always a force in the direction of motion. (incorrect student model)
Model 3:   Null model.

In the FCI , five questions activate models associated with the force – motion concept (questions 5, 9, 18, 22, 28).[15] As an example, consider question 5. (See figure 3.) The distracters "a", "b", and "c" represent three different responses associated with the same incorrect student model (Model 2). All of the three choices involve a force in the direction of motion. If a student selects one of these three choices, we consider that the student is using Model 2. (Here we use a model assignment scheme based on the student response to a single item. More complex situations can be considered. See [Bao 1999].) To use this method, we have to assume that if a student is cued into using a particular model, the probability for the student to apply the model inappropriately is small (<10%). Such probabilities can often be evaluated with interviews.[16] With this method, if a student answers "d" on this question, we assume that it is very likely for this student to have a correct model.[17] Choice "e" reflects the Aristotelian idea and is rarely held by students in our introductory physics class. If a student does choose this option, we consider this student as having a null model.

---

[13] Traditionally, when a bra and a ket are put together, they are displayed with the two vertical bars combined into a single bar, signifying the combination of the two elements into a single number. (Dirac, 1930.)

[14] For some tools to help design effective distracters and to see how different design may affect the measurement, please see (Bao and Redish, 2001) and (Bao 1999).

[15] In the FCI, two clusters of questions, those on Force-Motion and Newton III, provide most of the FCI's discriminatory power. For details on how we identified these questions using a quantitative argument, see (Bao and Redish 2001).

*Table 1. Associations between the physical models and the choices of the five FCI questions on Force-Motion concept.*

| Questions | Model 1 | Model 2 | Model 3 |
|---|---|---|---|
| 5 | d | a, b, c | e |
| 9 | a, d | b, c | e |
| 18 | b | a, e | c, d |
| 22 | a, d | b, c, e | |
| 28 | c | a, d, e | b |

We assume that there are clear associations between the three models and the responses corresponding to the five FCI questions in the force-motion cluster as listed in table 1. Notice that the mappings between model and item do not have to be one-to-one. Further, note that having the correct model does not imply having the correct answer. The student might have a correct model but employ it incorrectly. If there are known to be common errors in applying a correct model, we might want to include some of these errors as distracters. A good understanding of the most common student errors allows the construction of questions that probe both student model choice and student accuracy. In this analysis we only consider the students' model choice. This underlines the fact that a model analysis provides different information about student thinking than does a right/wrong analysis.

> 5. A boy throws a steel ball straight up. Consider the motion of the ball only after it has left the boy's hand but before it touches the ground, and assume that forces exerted by the air are negligible. For these conditions, the force(s) acting on the ball is (are):
>
> (A) a downward force of gravity along with a steadily decreasing upward force.
>
> (B) a steadily decreasing upward force from the moment it leaves the boy's hand until it reaches its highest point; on the way down there is a steadily increasing downward force of gravity as the object gets closer to the earth.
>
> (C) an almost constant downward force of gravity along with an upward force that steadily decreases until the ball reaches its highest point; on the way down there is only a constant downward force of gravity.
>
> (D) an almost constant downward force of gravity only.
>
> (E) none of the above. The ball falls back to ground because of its natural tendency to rest on the surface of the earth.

*Figure 3. Question 5 of FCI test*

---

[16] More detailed analysis on the certainty of such a model assignment is discussed in chapter 4 of reference 6. The probability can also be measured with specially designed questions where we can give a cluster of questions based on similar context settings with the leading ones (often simple) to test if the students are triggered into a particular model state and the following ones (somewhat more complex) to test if the students can apply the models correctly. We have also developed modeling schemes based on the student responses patterns on a series of questions. Details can also be found in reference 6.

[17] Note that this is not always the case. With some questions, students can choose the right answer for the wrong reasons. To obtain more accurate representations of student reasoning using our method, the wording of such items needs to be improved and the probability of student model crossover estimated through interviewing (Bao, 1999).

Using table 1, we can obtain an estimation of individual students' model states from students' responses. For example, if a student answers the five questions with "a", "d", "a", "d" and "b", the student probability vector is $(221)^T/5$. Using eq. (4), the model state for this student is $(\sqrt{2} \quad \sqrt{2} \quad 1)^T/\sqrt{5}$.

## *The Density Matrix: Model Estimation*

As discussed above, for a particular physical concept, a single student can have a consistent model (not necessarily a correct one), which is used for all questions related to the concept, a mixed model state where the student uses several models (correct and incorrect ones) inconsistently, or no clear model (no systematic logical reasoning involved in generating the response).

For a class probed by a given instrument, each student has an individual model state. Analysis using scores alone often fails to provide useful details on the students' real understanding of the physics concept (except in the case when most students consistently give correct answers). For example, a low score can be caused by a consistent incorrect model, calculation errors generated while using a correct model, random guessing, or a persistently triggered incorrect model for a student in a mixed model state. These different situations reflect important information on student understanding of physics; but they cannot be distinguished using an analysis based solely on scores. We introduce here a procedure we call *model estimation* that can provide a way to extract such information.

Using a group of questions associated with a single physics concept, we can measure and represent the single student model state with eq. (4). In the following, we use the example of the force-motion models and the FCI to demonstrate the model estimation algorithm. The FCI has 5 force-motion questions and involves three models, so $m = 5$ and $w = 3$. We can rewrite eq. (4) as:

$$|\mathbf{u}_k\rangle = \frac{1}{\sqrt{5}} \begin{pmatrix} \sqrt{n_1^k} \\ \sqrt{n_2^k} \\ \sqrt{n_3^k} \end{pmatrix} \quad (6)$$

where $n_i^k$ is the number of questions the $k^{th}$ student answered using the $i^{th}$ model.

We define *the single student model density matrix* for the $k^{th}$ student as ($w = 3$):

$$\mathcal{D}_k = |\mathbf{u}_k\rangle\langle\mathbf{u}_k| = \{\rho_{\eta\mu}^k\} = \frac{1}{m} \begin{bmatrix} n_1^k & \sqrt{n_1^k n_2^k} & \sqrt{n_1^k n_3^k} \\ \sqrt{n_2^k n_1^k} & n_2^k & \sqrt{n_2^k n_3^k} \\ \sqrt{n_3^k n_1^k} & \sqrt{n_3^k n_2^k} & n_3^k \end{bmatrix} \quad (7)$$

Although the single-student model density matrix clearly contains no more information about the student than does the model vector, (all the elements of the matrix are uniquely determined by the elements of the vector) the situation changes dramatically when we sum over all students in the class. We define the *class model density matrix* as the average of the individual students' model density matrices:

$$\mathcal{D} = \begin{bmatrix} \rho_{11} & \rho_{12} & \rho_{13} \\ \rho_{21} & \rho_{22} & \rho_{23} \\ \rho_{31} & \rho_{32} & \rho_{33} \end{bmatrix} = \frac{1}{N}\sum_{k=1}^{N}\mathcal{D}_k = \frac{1}{N}\sum_{k=1}^{N}\begin{bmatrix} \rho_{11}^k & \rho_{12}^k & \rho_{13}^k \\ \rho_{21}^k & \rho_{22}^k & \rho_{23}^k \\ \rho_{31}^k & \rho_{32}^k & \rho_{33}^k \end{bmatrix} \quad (8)$$

The class model density matrix retains important structural information about the individual student models which is otherwise lost if we only sum over the model vectors (this will produce the diagonal elements of the density matrix). By analyzing this matrix, we can study the features of the models used by the students in the class.

Now let's consider a population of students with diverse background. In solving a set of questions on a single concept, students in a class can be in a variety of situations on using their models. Three common situations are:

1. Most students in a class have the same model (not necessarily a correct one) and are self-consistent in using it.
2. The class population has several different models but each student only uses one model consistently. Thus the class of students can be partitioned into several groups each with a different but consistent model.
3. Students in the class can each have multiple physical models and use these models inconsistently, i.e., the individual students have mixed model states.

Note that these different situations are statistical features of the population which are intrinsically different from the probabilistic nature of individual student's model state. Corresponding to these different situations, the class model density matrix will show different structures (see figure 4). As indicated from eq. (8), the diagonal elements of the $\mathcal{D}$ reflect the percentage of the responses generated with the corresponding models used by the class. The off-diagonal elements reflect the consistency of the individual students' use of their models. Large off-diagonal elements indicate low consistency (large mixing) for <u>individual students</u> in their model use.

Using the class model density matrix, we can extract quantitative information on the distribution of student models for the class. One convenient method is to perform an eigenvalue decomposition to extract *class model vectors* (the eigenvectors of $\mathcal{D}$) and the eigenvalues.[18] A detailed discussion of the eigenvalue analysis is given in the appendix.

The analysis in the appendix demonstrates that the $\mu^{th}$ eigenvalue is the average of the squares of the overlap (dot product) between the $\mu^{th}$ eigenvector and the individual students' model vectors. Consequently, the eigenvalue is affected by both the similarity of the individual students' model vectors and the number of students with similar model state vectors. Thus if we obtain a large eigenvalue from a class model density matrix, it implies that many students in the class have similar single student model state vectors (the class has a consistent population). On the other hand, if we obtain several small eigenvalues, it indicates that students in the class behave differently from one another. Therefore, we can use the magnitude of the eigenvalues to evaluate the consistency of a class' population and the applicability of the simple form of the model analysis method.

$$\begin{bmatrix} 1 & 0 & 0 \\ 0 & 0 & 0 \\ 0 & 0 & 0 \end{bmatrix} \quad \begin{bmatrix} 0.5 & 0 & 0 \\ 0 & 0.3 & 0 \\ 0 & 0 & 0.2 \end{bmatrix} \quad \begin{bmatrix} 0.5 & 0.2 & 0.1 \\ 0.2 & 0.3 & 0.1 \\ 0.1 & 0.1 & 0.2 \end{bmatrix}$$

(a)            (b)            (c)
Consistent      Consistent      Inconsistent
one-model      three-model      three-model

*Figure 4. Examples of student class model density matrix: (a) an extreme case corresponding to the first type of class model condition where everyone has the same physical model (model 1); (b) the second type of class model condition where the class consists of three different groups of students each with a consistent physical model; (c) the third type of class model condition where many students have multiple physical models and are inconsistent in using these models.*

Using an eigenvalue decomposition to analyze the class model density matrix, we can obtain a quantitative assessment on the structure and the popularity of the students' common model states. We can evaluate two types of consistency: the consistency of individual students using different models, which is reflected by the structures of the class model states (mixed or pure), and the consistency among different students which is represented by the eigenvalues.

---

[18] In calculating the eigenvalues and eigenvectors, an algorithm called singular value decomposition (SVD) is often used. This algorithm is more stable than eigenvalue decomposition, although when the matrix is full rank and has non-zero eigenvalues, both methods give the same results.

As indicated from eqs. 14 and 15, if there is an eigenvector with a large eigenvalue, it contains the dominant features of the single student model vectors. We refer to this as a *primary* eigenvector. The additional eigenvectors act as corrections of less popular features that are not represented by the primary state. When considering the class as a single unit, a primary eigenvector gives good evaluation of the overall model structure of the class. However, if we regard the class as a composition of individual students, there can exists interesting details that cannot be extracted with a simple eigenvalue decomposition due to the fact that the eigenvalues method necessarily yields orthogonal eigenvectors.[19]

For example, suppose we have a class that can be divided into several groups of students, where students in each group all have similar model states and students from different groups have significantly different model states. In this situation an eigenvalue decomposition can give good results for the following two cases:

1. When the model states from different groups are nearly orthogonal,[20] the eigenvalue decomposition will produce eigenvectors that are similar to these model states.
2. When one of these student groups has a dominant population, the eigenvalue decomposition will produce a primary vector, with a large eigenvalue, very close to the model state held by this dominant group.

In the case when students are different but not "so" different (with a distribution of different but non-orthogonal model states), an eigenanalysis will not give appropriate model states. Rather, it will provide a set of orthogonal model vectors representing unique features of all the average students' model states. In the case that the eigenvalues are small, a scatter plot of the individual students' eigenvectors can suggest whether it might be useful to perform a cluster analysis, separating the class into distinct populations and determining the characteristics of those populations. In general, when the eigenvalue of a primary eigenvector is less than 0.65 and the student model states are mixed, it often indicates that the students in the class have a somewhat "flat" distribution of non-orthogonal model states. In such cases, plotting the angular distribution of the individual students' model states and/or conducting cluster analysis may provide more details on the population. However, when both eigenvalues and eigenvectors are considered, they can still provide a simple indicative evaluation on the population. In the example reported below, the primary eigenvalue is close to 0.8, which indicates that most students have similar model states.

### *Representing the Class Model State – The Model Plot*

In many situations we have encountered, students often have two dominant models, a correct one and a common misconception. To conveniently represent and study the states and changes of student models in this situation, we construct a two-dimensional graph or *model plot* to represent the student usage of the two models. For example, suppose we study the first two models in a 3-model situation. A class model state, $\mathbf{v}_\mu = (v_{1\mu}, v_{2\mu}, v_{3\mu})^T$, can be represented as a point in a two-dimensional space in which the two axes represent the probabilities that a student in the class will use the corresponding models on one of the items of the probe instrument. The state is represented by a point (point B in figure 5) that we refer to as the *class model point* on a plot with $P_1 = \sigma_\mu^2 v_{1\mu}^2$ as the vertical component and $P_2 = \sigma_\mu^2 v_{2\mu}^2$ as the horizontal component.

When the eigenvalue of a class model state is small, the class model point will be close to the origin. On the other hand, a state with a large eigenvalue will be close to the line going through (0, 1) and (1, 0), which is the upper boundary of the allowed region of the model plot.[21] In the case when a class model state vector has small elements on model dimensions that are not considered ($v_{3\mu}$ in this case), which often occurs, we can make an approximation letting $\sigma_\mu^2(1-v_{3\mu}^2) \cong \sigma_\mu^2$. Then the distance between a model point and the upper boundary can be used to estimate the eigenvalue of the corresponding model state. Defining *d* as the distance between a model point and the upper boundary, this estimation can be calculated with:

$$\sigma^2 \cong \left(1 - \frac{d}{\frac{1}{\sqrt{2}}}\right) = \left(1 - d \cdot \sqrt{2}\right) \qquad (9)$$

---

[19] It is also a general problem that will be encountered when attempting to represent the distributive results of a population with several definite items (vectors, values, etc.).

[20] This limits the number of such groups to be equal to the dimensions of the related model space.

[21] Since the two coordinates represent probabilities and the sum of the probabilities must be less than or equal to 1, a class point must lie below the line P1+P2=1. In addition, since each probability must be positive, the class point must lie within the triangle bounded by the points (0,0), (1,0), and (0,1).

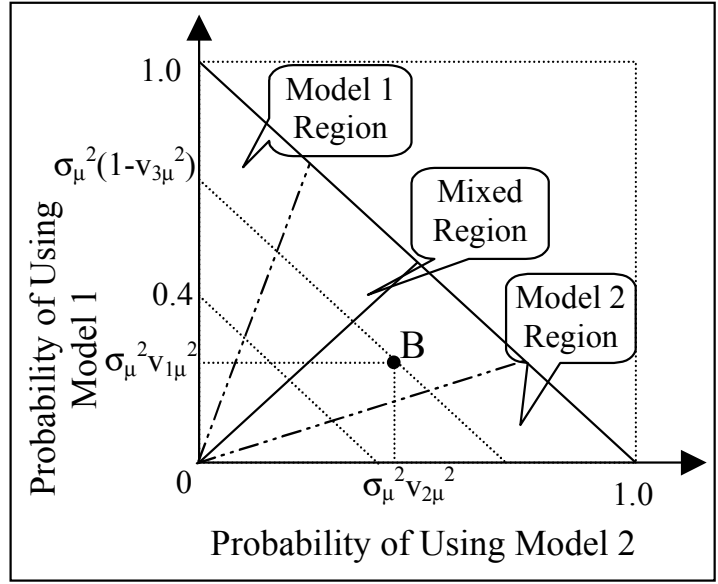

Figure 5. *Model regions on model plot. Model 1 (Model 2) region represents comparatively consistent model states with dominant model 1 (model 2) components. Mixed model region represents mixed model states.*

### *Describing Model Mixing Features*

When analyzing student model structures in a case where the model space is dominated by two models we can represent the student model states on a 2-D model plot as shown in figure 5. In order to describe the different regions of the plot, we separate the plot by drawing two straight lines from the origin with slopes equal to 1/3 and 3 respectively (see figure 5). We also draw the line corresponding to the condition $P_1+P_2 = 0.4$. With these lines, we partition the model plot into four regions: the *model 1 region*, the *model 2 region*, the *mixed region*, and the *secondary model region* (model states with eigenvalues smaller than 0.4) as shown in figure 5. In a class that has a primary model point in the model 1 region (or model 2 region), the individual students have comparatively consistent model states with dominant model 1 (or model 2). In a class with a primary model point in the mixed region, the individual students have predominantly mixed model states, i.e., most of the individual students are inconsistent in using the different physical models. The secondary model region represents model states with small eigenvalues, which reflect less popular features of the class behavior. In most cases we have studied, there is one primary model state with an eigenvalue 3 to 4 times larger than the second largest eigenvalue. In these cases, the primary model state alone provides a good overview of the class's model state.

The model plot can visually present much information about the student model states on the same graph (e.g. the consistency of the population, the consistency of individual students, and the types of models used my individual students). We can also put the pre and post model states from different classes together on the same plot, making it much easier to see the patterns and shifts of the different classes' model states.

## V. Model Analysis of FCI Data

Using the model estimation method, we analyzed FCI data from the pre-post testing of 14 introductory mechanics classes (Physics 161) at the University of Maryland.[22] The students are mostly engineering majors. All the classes had traditional lectures three hours per week and were assigned weekly readings and homework consisting of traditional textbook problems. All of the students also had one hour per week of small group (N~30) teaching assistant (TA) led recitations. In half of the classes recitations were traditional TA-led problem solving sessions (students asking questions and the TA modeling solutions on the board). The other half received recitations taught with *Tutorials* (McDermott & Shaffer, 1998). These sessions consisted of students working together in groups of 3-5 on research-based guided-discovery worksheets. The worksheets often used a cognitive conflict model and helped students develop qualitative reasoning about fundamental physics concepts. In the following analysis, we use the five FCI questions on the force-motion concept as an example to demonstrate the model estimation algorithm.

Using the item-based modeling scheme in table 1 and following the procedures from eqs. (2) to (8), we calculated the average student initial model state on force-motion by combining all classes (778 students). The results are shown in table 2. As we can see from table 2, the eigenvalues for the class states corresponding to the null models are very small. This indicates that most students use either the correct expert model or the incorrect naïve model and the model space defined from the qualitative research matches well with this population. In addition, the primary class model states (state with the largest eigenvalue) of all classes have eigenvalues around 0.8. Therefore, the primary state alone can give a fairly good description of the class. Using the results in table 2, the student class model states on the force–motion concept are displayed on a model plot spanned by model 1 (expert model) and model 2 (naïve model) (see figure 6). For each type of class, we plot the class' primary model state. The initial states of both types of classes are nearly the same and can be interpreted as that before instruction most students in the two classes consistently use the incorrect model on all the questions related to force-motion.

*Table 2. Results of class model density matrices and class model states on Force-Motion concept with data from UMd students.*

| Force – Motion | Tutorial | | | | | |
|---|---|---|---|---|---|---|
| | Pre | | | Post | | |
| Density Matrix | $\begin{bmatrix} 0.27 & 0.23 & 0.02 \\ 0.23 & 0.69 & 0.07 \\ 0.02 & 0.07 & 0.04 \end{bmatrix}$ | | | $\begin{bmatrix} 0.66 & 0.28 & 0.03 \\ 0.28 & 0.31 & 0.02 \\ 0.03 & 0.02 & 0.03 \end{bmatrix}$ | | |
| Eigenvalues | 0.80 | 0.17 | 0.03 | 0.82 | 0.15 | 0.03 |
| Eigenvectors | $\begin{pmatrix} 0.40 \\ 0.91 \\ 0.09 \end{pmatrix}$ | $\begin{pmatrix} -0.92 \\ 0.39 \\ 0.07 \end{pmatrix}$ | $\begin{pmatrix} 0.03 \\ -0.12 \\ 0.99 \end{pmatrix}$ | $\begin{pmatrix} 0.87 \\ 0.48 \\ 0.05 \end{pmatrix}$ | $\begin{pmatrix} -0.49 \\ 0.87 \\ 0.02 \end{pmatrix}$ | $\begin{pmatrix} 0.03 \\ 0.04 \\ -0.99 \end{pmatrix}$ |
| Force – Motion | Traditional | | | | | |
| | Pre | | | Post | | |
| Density Matrix | $\begin{bmatrix} 0.27 & 0.22 & 0.03 \\ 0.22 & 0.68 & 0.08 \\ 0.03 & 0.08 & 0.05 \end{bmatrix}$ | | | $\begin{bmatrix} 0.46 & 0.25 & 0.03 \\ 0.25 & 0.50 & 0.05 \\ 0.03 & 0.05 & 0.04 \end{bmatrix}$ | | |
| Eigenvalues | 0.79 | 0.17 | 0.04 | 0.74 | 0.23 | 0.03 |
| Eigenvectors | $\begin{pmatrix} 0.40 \\ 0.91 \\ 0.12 \end{pmatrix}$ | $\begin{pmatrix} -0.92 \\ 0.39 \\ 0.03 \end{pmatrix}$ | $\begin{pmatrix} 0.02 \\ 0.12 \\ -0.99 \end{pmatrix}$ | $\begin{pmatrix} 0.67 \\ 0.73 \\ 0.08 \end{pmatrix}$ | $\begin{pmatrix} -0.74 \\ 0.67 \\ 0.06 \end{pmatrix}$ | $\begin{pmatrix} 0.01 \\ 0.10 \\ -0.99 \end{pmatrix}$ |

---

[22] Data collected by J. M. Saul at the University of Maryland (UMd).

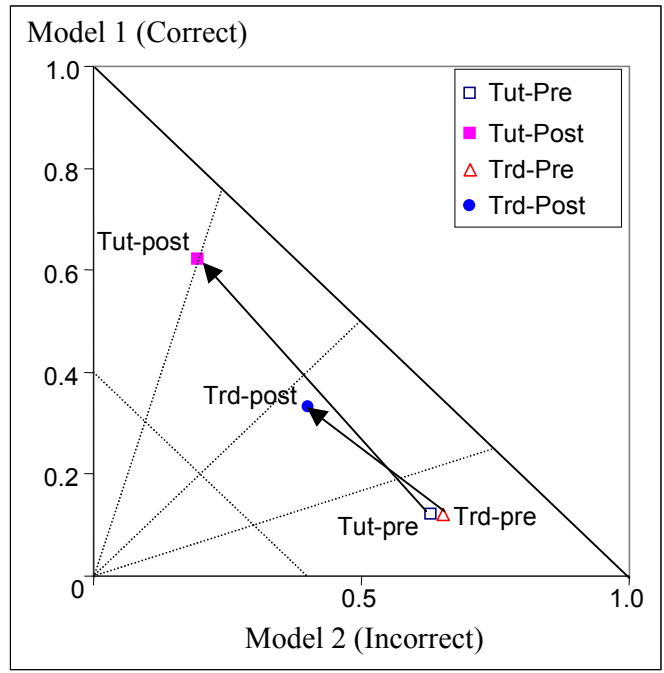

*Figure 6. Model plot of student class model states on Force-Motion with FCI data from University of Maryland. For each type of class, we plotted, for pre and post results, the first two class model states (states with first and second largest eigenvalues). The two arrows represent the shifts of the first model states for pre and post results of Tutorial and traditional classes.*

After instruction, the model states of the tutorial classes still indicate that most students use the correct model rather consistently. On the other hand, the primary model state of the traditional classes indicates a mixed model state, which shows that most students in the class are inconsistent in using their models. Since the model state is nearly a perfect mix (half to half), a particular student is likely to use the correct model on half of the questions and use the incorrect model to the other half of the questions. This result provides a piece of evidence that validates our treatment of context dependence with conditioned random process in considering a population, and also indicates that the five FCI questions on force-motion are well designed and are appropriate for assessment of students' conceptual knowledge concerning context-dependence.

## VI. Comparing Model Analysis and Factor Analysis

Factor analysis is a popular tool in education research. It can give the correlation (consistency) of student behavior on different test items; however, if the results are based solely on student scores (i.e. whether they have the correct responses to a set of questions or not), it can fail to provide complete information on students' states of conceptual understanding. In a cluster designed to probe knowledge of a particular concept, even if all students in a class are in similar mixed model states, they will tend to answer some items correctly and others incorrectly in random ways. As a result, factor analysis will not be able to identify a strong factor and the dominant common element in the class's knowledge state will be missed (Huffman and Heller, 1995).

Unlike factor analysis, which usually tries to draw the components of student reasoning directly from test data, Model Analysis puts information from qualitative research into the analysis to determine the state space. Thus, model analysis is able to provide a set of tools that can be used to investigate different possibilities for the students' states of understanding and how those states are related to accepted scientific models. Consider an idealized example to demonstrate how these two methods differ. Suppose we give four multiple-choice questions to a class of 100 students ($m = 4$, $N = 100$). All four questions probe understanding of a single physics concept that might activate one of two models -- model A and model B ($w = 2$). Consider two situations:

Case 1: All students in the class are self-consistent. Half of them use model A on all four questions, and the other half use model B on all four questions.

Case 2: All students are equally mixed between model A and model B: All students' applications of the models are context dependent. They use model A and model B equally, so each student applies model A to two questions and model B to the other two but the choices of which questions correspond to which model is random.

In case 1, the results from both methods are calculated in table 3. As we can see, the results from model analysis show two states with equal weights indicating that the class has two groups each of which consistently use one of the models. The results from factor analysis give a single factor, which shows that all the students are consistent. The result from factor analysis does not tell in which way the students are being consistent. (This can, however, be supplemented by the information about the class scores.)

*Table 3. Results from model analysis and factor analysis for a class having two equal populations each with a consistent model.*

| Model Analysis | | Factor Analysis | |
|---|---|---|---|
| Density Matrix | $\frac{1}{2}\begin{bmatrix} 1 & 0 \\ 0 & 1 \end{bmatrix}$ | Correlation Matrix | $\begin{bmatrix} 1 & 1 & 1 & 1 \\ 1 & 1 & 1 & 1 \\ 1 & 1 & 1 & 1 \\ 1 & 1 & 1 & 1 \end{bmatrix}$ |
| Eigenvalues | $\sigma_1^2 = \frac{1}{2}, \sigma_2^2 = \frac{1}{2}$ | Eigenvalues | $\sigma_1^2 = 4, \sigma_j^2 = 0, (j = 2,3,4)$ |
| Class Model States | $\begin{pmatrix} 1 \\ 0 \end{pmatrix}, \begin{pmatrix} 0 \\ 1 \end{pmatrix}$ | Factors | $\frac{1}{2}\begin{pmatrix} 1 \\ 1 \\ 1 \\ 1 \end{pmatrix}$ |

*Table 8. Results from model analysis and factor analysis for a class having a single population with an equally mixed model state.*

| Model Analysis | | Factor Analysis | |
|---|---|---|---|
| Density Matrix | $\frac{1}{2}\begin{bmatrix} 1 & 1 \\ 1 & 1 \end{bmatrix}$ | Correlation Matrix | $\begin{bmatrix} 1 & -0.33 & -0.33 & -0.33 \\ -0.33 & 1 & -0.33 & -0.33 \\ -0.33 & -0.33 & 1 & -0.33 \\ -0.33 & -0.33 & -0.33 & 1 \end{bmatrix}$ |
| Eigenvalues | $\sigma_1^2 = 1, \sigma_2^2 = 0$ | Eigenvalues | $\sigma_1^2 = 1.33, \sigma_2^2 = 1.33,$ $\sigma_3^2 = 1.33, \sigma_4^2 = 0$ |
| Class Model States | $\frac{1}{\sqrt{2}}\begin{pmatrix} 1 \\ 1 \end{pmatrix}$ | Factors | $\frac{1}{2}\begin{pmatrix} 1 \\ -1 \\ -1 \\ 1 \end{pmatrix}, \frac{1}{\sqrt{2}}\begin{pmatrix} 0 \\ 1 \\ -1 \\ 0 \end{pmatrix}, \frac{1}{\sqrt{2}}\begin{pmatrix} -1 \\ 0 \\ 0 \\ 1 \end{pmatrix}, \frac{1}{2}\begin{pmatrix} 1 \\ 1 \\ 1 \\ 1 \end{pmatrix}$ |

The results for case 2 are displayed in table 4. For both methods, eigenvectors with eigenvalues equal to zero are omitted. Since the students are assumed to be equally mixed with model A and model B the probability for a single student to use either model A or model B is equal for all questions. As we can see, the results from model analysis indicate a single perfectly mixed class model state with 100% dominance ($\sigma^2=1$). On the other hand, since the students are inconsistent in answering the questions, factor analysis gives insignificant random-like correlation between different questions and show no dominant factors. Such situation is often interpreted as if there is no factor in the data. This example shows that score-based factor analysis does not deal adequately with the randomness in individual students' use of conceptual models.

## VIII. Summary

In this paper, we have introduced Model Analysis, a method to analyze the content of student knowledge. It begins with the cognitive assumption that students are often inconsistent in their use of mental models in situations that an expert would consider equivalent. We suggest that the best way to treat this situation for the evaluation of goal-oriented instruction is by considering the student as being able to simultaneously be in different model states with probabilities for the activation of different states. Model Analysis allows the assessment on the probabilities of students' use of these alternative models. The results can be used to analyze student understanding and/or the features of the different instruments.

Model Analysis presents a way to integrate the qualitative knowledge gained from student interviews with the quantitative analysis of multiple-choice instruments. Use of this method works as follows:

1. Through systematic research and detailed student interviews, common student models are identified and validated so that these models are reliable for a large population of students with a similar background.

2. This knowledge is then used in the design of a multiple-choice instrument. The distracters are designed to activate the common student models and the effectiveness of the questions is validated through research.

3. To use Model Analysis, one then classifies the students' responses by model and creates a state representing the student probability amplitude in the model space.

4. The individual student model states are used to create a density matrix, which is then summed over the class. The eigenvalues and eigenvectors of the class density matrix give information about the state of the class's knowledge.

In constructing a measurement of student conceptual understanding, there is often a "communication" problem – students can use the same terminology (or a statement) as used by an expert but with a different understanding. Simply using a word or a statement often fails to extract the actual underlying reasoning, which usually can only be obtained by analyzing how students apply their knowledge. Model Analysis relies heavily on qualitative methods. By conducting systematic research, it is expected that the identified student models reflect the majority of different types of student understandings, and the multiple-choice instruments do not contain significant communication problems – the distracters are designed to reflect not a simple usage of a word or statement but rather the results of students' application of their models. That is, we use interviews to identify the students' actual reasoning common to a large population and use research-based multiple-choice instruments, with the algorithms in Model Analysis, to measure the students' use of these popular types of reasoning in learning. The combination of the two methods can partially solve the communication problem and yet provide an effective and stable tool to probe large classes. Once a reliable package is developed, instructors with limited training can easily implement Model Analysis instruments to obtain immediate feedback from students and comparatively rich information on the students' actual understandings.

It is often argued that by putting in our knowledge of the responses of some students we limit the framework for students' possible models. In Model Analysis, we always include a null model space to include possibilities that may be missing when the test is designed. In early stages of research, Model Analysis could be used with open-ended questions and the results classified by common models using phenomenography (Marton, 1986). If a large null model element is identified in the analysis, it immediately alerts us that the population being tested may have possible models that are not understood, suggesting the need for further research. Therefore, model analysis provides a set of tools that can be used to evaluate the features of the instruments and therefore, can be used in a systematic cyclic process of research and development.

The results from Model Analysis give more explicit information on improving instruction than score-based analysis. With the knowledge of students' model states and changes of such states, instructors can see more directly about the possible causes of the student difficulties and develop better instruction strategies to help the student.

## Appendix: Eigenvalue analysis of $\mathcal{D}$

Since $\mathcal{D}$ is symmetric and all data are real and nonnegative definite, the eigenvalues are all real nonnegative numbers, which we denoted by $\sigma^2_1, \sigma^2_2, \ldots \sigma^2_w$. Define the eigenvectors of $\mathcal{D}$ as $\mathbf{v}_\mu$ (a column vector) where $\mu = 1, \ldots, w$ are the indices for different class model states. Then the matrix of eigenvectors can be written as

$$\mathbf{V} = [\mathbf{v}_1, \ldots, \mathbf{v}_\mu, \ldots, \mathbf{v}_w]$$

This matrix transforms $\mathcal{D}$ into a diagonal form. For a 3-D model space we can write

$$\mathbf{V}^T \mathcal{D} \mathbf{V} = [\Sigma^2]^{w=3} = \begin{bmatrix} \sigma_1^2 & 0 & 0 \\ 0 & \sigma_2^2 & 0 \\ 0 & 0 & \sigma_3^2 \end{bmatrix}$$

Now let us see how the information of the individual students' model states are stored in $\mathcal{D}$ and what can be learned from the eigenvalues and eigenvectors. Consider a class with $N$ students. The class model density matrix can be written as

$$\mathcal{D} = \frac{1}{N} \sum_{k=1}^{N} \mathcal{D}_k = \frac{1}{N} \sum_{k=1}^{N} |\mathbf{u}_k\rangle\langle\mathbf{u}_k| \tag{10}$$

Using the eigenvectors and eigenvalues, $\mathcal{D}$ can also be written as

$$\mathcal{D} = \sum_{\mu=1}^{w} \sigma_\mu^2 \cdot |\mathbf{v}_\mu\rangle\langle\mathbf{v}_\mu| \tag{11}$$

Applying this to an eigenvector, we can write:

$$\mathcal{D}|\mathbf{v}_\mu\rangle = \frac{1}{N} \sum_{k=1}^{N} |\mathbf{u}_k\rangle\langle\mathbf{u}_k|\mathbf{v}_\mu\rangle = \sigma_\mu^2 \cdot |\mathbf{v}_\mu\rangle \tag{12}$$

Define $a_{\mu k}$ as the agreement between the $k^{th}$ student's model vector $\mathbf{u}_k$ and the $\mu^{th}$ eigenvector of $\mathcal{D}$. We have:

$$a_{\mu k} = \langle\mathbf{u}_k|\mathbf{v}_\mu\rangle = \langle\mathbf{v}_\mu|\mathbf{u}_k\rangle \tag{13}$$

Then eq. (13) can be rewritten as

$$\mathcal{D}|\mathbf{v}_\mu\rangle = \frac{1}{N} \sum_{k=1}^{N} |\mathbf{u}_k\rangle\langle\mathbf{u}_k|\mathbf{v}_\mu\rangle = \frac{1}{N} \sum_{k=1}^{N} a_{\mu k} \cdot |\mathbf{u}_k\rangle = \sigma_\mu^2 \cdot |\mathbf{v}_\mu\rangle$$

which yields

$$|\mathbf{v}_\mu\rangle = \frac{1}{\sigma_\mu^2 \cdot N} \sum_{k=1}^{N} a_{\mu k} \cdot |\mathbf{u}_k\rangle \tag{14}$$

Thus an eigenvector of $\mathcal{D}$ is a weighted average of all the individual student model vectors with weights equal to the agreements between the eigenvector and the single student model vectors. Therefore, the class model states represented by these eigenvectors are the set of states that reflect the salient features of all the individual student model vectors.

If we left multiply eq. 14 by its conjugate, we then have

$$\langle \mathbf{v}_\mu | \mathbf{v}_\mu \rangle = \frac{1}{\sigma_\mu^2 \cdot N} \sum_{k=1}^{N} a_{\mu k} \cdot \langle \mathbf{v}_\mu | \mathbf{u}_k \rangle = \frac{1}{\sigma_\mu^2 \cdot N} \sum_{k=1}^{N} a_{\mu k}^2 = 1$$

and

$$\sigma_\mu^2 = \frac{1}{N} \sum_{k=1}^{N} a_{\mu k}^2 \qquad (15)$$

This result indicates that the $\mu^{th}$ eigenvalue is the average of the squares of the agreements between the $\mu^{th}$ eigenvector and the individual students' model vectors. Consequently, the eigenvalue is affected by both the similarity of the individual students' model vectors and the number of students with similar model state vectors. Thus if we obtain a large eigenvalue from a class model density matrix, it implies that many students in the class have similar single student model state vectors (the class has a consistent population). On the other hand, if we obtain several small eigenvalues, it indicates that students in the class behave rather differently from one another. Therefore, we can use the magnitude of the eigenvalues to evaluate the consistency of a class' population.